**Assessing technical and cost efficiency of research activities: A case study of the Italian university system**[1]

Giovanni Abramo[a,b,*], Ciriaco Andrea D'Angelo[a]

[a] *Laboratory for Studies of Research and Technology Transfer, University of Rome "Tor Vergata", Italy*

[b] *Italian Research Council*

## Abstract

This paper employs data envelopment analysis (DEA) to assess both technical and cost efficiency of research activities of the Italian university system. Differently from both peer review and the top-down discipline-invariant bibliographic approaches used elsewhere, a bottom-up bibliometric methodology is applied. Publications are assigned first to authors and then to one of nine scientific and technical university disciplinary areas. Inputs are specified in terms of the numbers of full, associate and assistant professors and outputs as the number of publications, contributions to publications and their scientific impact as variously measured across the disciplines included. DEA is undertaken cross-sectionally using the averages of these inputs and outputs over the period 2001–2003. The results typically show much variation in the rankings of the disciplinary areas within and across universities, depending on the efficiency indicator employed.

* Corresponding author: Dipartimento di Ingegneria dell'Impresa, University of Rome "Tor Vergata", Via del Politecnico 1, 00133 Rome - ITALY, tel. and fax +39 06 72597362, abramo@disp.uniroma2.it

## 1. Introduction

Scholars, policy makers and the top managers of research institutions are increasingly interested in the evaluation of research systems. A growing number of initiatives have been launched at international levels to analyze the performance of national research systems and their individual research institutions. Some studies arise from the international liberalization of research markets, particularly in association with university training (SJTU, 2006; THES, 2006). Other work arises from the need for a common platform to compare and share practices and policies, especially for countries which aim at developing a knowledge-based economy (EU-DGR, 2005; NSF, 2006).

At a national level, research assessment exercises are mainly aimed at an efficient allocation of resources among research institutions and/or at stimulating increased levels of research productivity on the part of the funding recipients. At the level of single institutions, the implementation of evaluation systems is part of a "top-down" logic used to adapt to the resource allocation processes of national governments or is also aimed at promoting continuous improvement in those institutional contexts that are most exposed to market competition (Georghiou and Larédo, 2005).

This paper will focus on the approaches used to assess the efficiency of national public research systems and above all on methodological aspects which are particularly useful for creating decision support systems for efficient resource allocation.

The most common research evaluation methods fall into two general categories: peer review methodologies and the so-called bibliometric methodology. Peer-review methodologies are based on qualitative judgments by sectorial experts who are given a sample of research products from each institution under analysis. Such methodologies appear to suffer severe limitations (Moxham and Anderson, 1992; Horrobin, 1990), most of which can be traced back to constraints arising from subjectivity of the judgments. The time and implementation costs for peer review methodology are another deterring element. In large-scale assessments, such as those carried out at national levels, it may be too costly to evaluate all the research products of the institutions under consideration, and therefore to measure productivity. As a consequence, final assessments often refer solely to the most notable research products, as selected by the research organizations under evaluation.

The bibliometric approach is based on use of indicators which are linked to two basic drivers: the publication itself and any citations it has obtained over time. Publications and citations are usually surveyed by querying ad hoc databases, such as those developed by Thomson Reuters[1]. Unlike the peer-review approach, the availability of the output data (publications), as well as of input data (such as the human and financial resources employed) allows the measurement of production efficiency. The use of purely bibliometric indicators has been the subject of specific technical and methodological cautions (Van Raan, 2005). The main limitation, having the greatest impact on potential assessment and allocation purposes, arises from considering scientific publication alone as a proxy of research output, while overlooking all other forms in which new knowledge is codified. Nonetheless, this assumption does have a host of empirical evidence to warrant its use[2].

Both the peer-review and the bibliometric methodologies have advantages and disadvantages, which have been extensively discussed in the literature. Although no single vision or common approach has been developed thus far, the peer-review method seems to be preferred over the bibliometric one (a notable case being Great Britain's

Research Assessment Exercise[3]), yet the bibliometric assessment methodology offers several advantages over the peer-review approach: it is low-cost, non-invasive, easy to implement, ensures both rapid updates and inter-temporal comparisons, is based on objective qualitative-quantitative data, and offers a high degree of representativeness of the surveyed universe, thereby permitting measurements of productivity. The correct use of indicators of production based on publication databases is, however, subject to a number of technical limitations, of which two can be noted:

- The first limitation concerns problems involved in the attribution of publications, with their attendant citations, to the relevant authors' organizations. The attribution process is quite complex and is afflicted by a series of potentially damaging errors arising from a complex of causes, which have been thoroughly illustrated in literature (Georghiou and Larédo, 2005).
- The second limitation is the representativeness of the journals covered in the source databases. In the instance of Thomson Reuters' Web of Science (WoS) SCI, over 10,000 international journals are covered. This cannot be considered an exhaustive representation of the complex scientific publication universe. The representativeness of the journals covered also varies according to discipline, and coverage is clearly higher for technical-scientific areas than for arts and humanities.

Furthermore, as for the methodology, the techniques used to aggregate the data at the outset of the analysis can induce serious problems. This is due to the fact that the productivity of scientists in terms of numbers of publications varies with discipline. Thus, comparing organizations that deal in different disciplines without allowance for their heterogeneity can lead to incorrect conclusions. Various studies quantify the distortions due to such limits and warn against using bibliometric approaches without the necessary cautions (most notable are Abramo and D'Angelo, 2007; Van Raan, 2005; MacRoberts and MacRoberts, 1996; Moed, 2002).

In short, it appears that bibliometric analytical methods have technical and methodological limitations which prevent them from being used on a large scale. Moreover, the larger the scale of the assessment exercises, the greater impact the limitations will have. This explains in part why bibliometric approaches have had only marginal use for national assessments. Previous bibliometric studies aimed at assessing the research and teaching performances of universities (Abbott and Doucougalios, 2003 and Worthington and Lee, 2008, for the Australian university system; Flegg et al, 2004, and Athanassopoulos and Shale, 1997, for Great Britain; Baek, 2006, for the USA) dismiss the problem of varying prolificacy among scientific disciplines by assuming that, by virtue of their size, all universities will have a set of similar disciplines of scientific activity. Yet this assumption is highly unlikely to be correct.

In an attempt to assess the research productivity of Italian universities, CRUI (2002) rejected the above-mentioned assumption and adopted a "top-down" approach: SCI publications were assigned to disciplinary areas and the number of publications in each area was subsequently divided by the number of researchers employed in that particular area within each university[4]. However, this procedure also induced significant distortions (as indicated by Abramo et al., 2008), due to the fact that researchers may, with some frequency, publish their works in journals that fall within disciplines other than those in which the researcher is classified.

The only way to avoid these problematic assumptions and overcome the current limitations to bibliometric approaches is to take a "bottom-up" approach: to first identify each publication with its real authors, to measure then its relative impact as compared to those falling in the same WoS category[5], and only later to measure each author's relative performance as compared to those falling in the same scientific disciplinary sector to which the authors belong. This is the method adopted for this study, which covers all SCI™ (CD-rom version) articles and reviews by Italian scientists, published between 2001 to 2003, the same period covered by the first and only Italian research assessment exercise.

In addition to the limitations discussed above, current bibliometric approaches for research evaluation tend to be used solely for assessment of technical efficiency, of research institutions. To the authors' knowledge no examples of assessing allocative efficiency, and thus cost efficiency as a whole, have been published so far.

The precise objective of the present study is to develop a measurement methodology for technical and allocative efficiency of public research institutions based on the application of *data envelopment analysis* (DEA) to bibliometric data (a "bottom-up" approach); to implement and evaluate the approach in the study of Italian universities active in technical-scientific disciplines; and to compare the results with those deriving from the application of other indicators of productivity.

The remainder of this article describes the study undertaken by the authors. In particular, section 2 describes the methodology used: the domain of investigation and sources, the processes of data extraction and source-data post-codification used to survey the scientific production of Italian universities, and the non-parametric input/output model used for the assessment of research activities. Section 3 shows the results obtained from the analysis of technical, allocative and cost efficiencies by disciplinary areas within each university. Section 4 provides the scores obtained from a series of efficiency indicators applied to the Italian universities operating in each disciplinary area, analyzes performances of individual disciplines within the same university, and finally compares the results of the analysis with outcomes from application of other types of bibliometric assessment models. The authors' considerations conclude the article.

## 2. Methodology

The surveyed field includes all Italian universities carrying out research in technical-scientific *university disciplinary areas* (UDA)[6], for the survey period from 2001 to 2003. In these disciplinary areas the SCI™ database indexes practically all articles authored by Italian academics.

The methodology adopted to measure technical and cost efficiency is based on the application of the DEA technique to bibliometric data for the scientific production of the universities under consideration. The methodology is defined as "bibliometric non-parametric".

DEA has already been used in several documented cases (Charnes et al., 1978; Banker et al., 1984). In general, non-parametric techniques (unlike parametric techniques) do not require *a priori* definitions of closed models for the production function. As well, the DEA technique is unlike partial indicators that depend on the simple normalization of a research output value in relation to the size of an input value. DEA embodies a truer representation of the complexity of research systems, which produce multiple outputs from diverse inputs. It is clear that the advantages and simpler requirements for applying DEA have favored its use for evaluations of education activities of entire national university systems.

In the current study a “costs” vector containing the economic weights of a group of input factors is used to assess the cost efficiency (CE) of surveyed units (decision making units, or DMU) as a product of two factors: technical efficiency (TE) and allocative efficiency (AE). The calculations of the frontier and the efficiency indexes were carried out using the Data Envelopment Analysis Program (DEAP) developed by Tim Coelli of the Centre for Efficiency and Productivity Analysis, University of Queensland, Australia. The input and output variables represented in the DEA model are listed in Table 1.

| Variable | Type | Acronym |
|---|---|---|
| Number of full professors | Input | FP |
| Number of associate professors | Input | AP |
| Number of assistant professors | Input | RF |
| Number of publications | Output | PU |
| Contribution to publications | Output | PC |
| Scientific Strength | Output | SS |

***Table 1: Variables of the DEA survey model used for the analysis***

Input data in the model describe both scientists employed in technological-scientific areas and attendant costs. Data on tenured university staff for each teaching level were extracted from the database of the Ministry of Universities and Research as of the last calendar day of each of the years taken into consideration. The choice to distinguish three staff ranks was intended to permit analysis of the levels of production “quality” and costs for the different academic ranks. Analysis of the cost data for university staff employed over the three years under consideration[7] permitted the calculation of weights of input factors for each academic rank. The resulting three-dimension vector (1.814; 1.370; 1.000), respectively indicating full professor, associate professor and assistant professors relative costs[8], is invariant among the disciplinary areas. No information on capital inputs per university and disciplinary area was available. However, human resources’ wages represent around 90% of overall research costs of Italian universities.

The output variables used to feed the DEA model describe three assessment dimensions of scientific production, namely: quantity of publications (PU), contribution by the author (PC), and impact (SS):

- The PU output, for the $i^{th}$ disciplinary area of the $j^{th}$ university, is calculated as the sum of publications in area *i* having at least one author from university *j*.
- The PC index is similar to PU but takes into account the authors’ “contribution”. It is measured as:

  $$PC_{ij} = \sum_{publications} \frac{b_{ij}}{c}$$

  where $b_{ij}$ equals the sum of the number of authors of the publication belonging to the $i^{th}$ disciplinary area of the $j^{th}$ University, and *c* is the total number of authors of the publication.
- The scientific strength, SS equals the weighted average of total publications (PU) by each university within each disciplinary area. The weights used for this study refer to the impact factor of the journal in which each article is published[9]. The scientific strength of university *j* in the disciplinary area *i* is measured as:

  $$SS_{ij} = \sum_{publications} \overline{IF}_{ij}$$

where $\overline{IF}_{ij}$ equals the normalized impact factor[10] of each publication belonging to the $i^{th}$ disciplinary area of the $j^{th}$ University.

Both input and output data inserted in the DEA model are the average values over the 3 year period under consideration. This model, although simplified, quite faithfully represents the research production function in Italian universities. It should be noted that Italian universities all have a similar "public" structure, highly similar management systems and a completely homogenous legal framework[11].

Table 2 shows the average input-output data for the Italian university system over the surveyed years. Data on output variables were extracted from a database, Observatory of Public Research (Orp), specifically developed for the purpose by the authors. The database collects and sorts information on the scientific production of scientists from Italian universities (a bottom-up approach). It enables aggregation operations at the higher levels of discipline, school and university with better degrees of accuracy than extractions that are simply aggregated by university (a top-down approach). The database was developed by extracting all the publications from the CD-rom version of SCI™ (articles and reviews only) that included an author from at least one Italian university. To do this it was necessary to identify and reconcile all the possible variations of name forms for individual Italian universities used in the "addresses" field of the SCI™. These publications amount to a total of 62,523 for the period from 2001 to 2003.

Subsequently, the author names in the list of the extracted SCI™ publications were disambiguated on the basis of the lists of university tenure teaching staff in place on the last day of every year under consideration. Such procedure proved particularly taxing, first because the records in the SCI™ show no link between the "author list" and the "address list" of a publication, and secondly because of particularly strong homonymy, which, in turn, results from two distinct factors: the coding of names in the "author list" of SCI™ (last name and initial of first name), and the size of the observed population (over 36,000 university research staff).

Therefore, a specific disambiguation algorithm was formulated to retrace the "publication-author-university-disciplinary area" links. A total of 9,298 publications were discarded from those originally listed because the authors, although they indicated an academic affiliation, were not included in the official list of Italian academic staff. A series of tests carried out on disambiguated publications showed a very low percentage (around 3%) of "false authorships" (i.e. publications with attribution errors). Some limitations due to errors in the original SCI™ data source still remain, as well as those caused by authors who have provided an incorrect indication of their name or institutional affiliation. However, given the domain of this study and the limited number of errors found during the analysis, it is quite reasonable to think that such errors and limitations do not favor particular universities over others and that they can therefore be regarded as statistically acceptable.

After being disambiguated, the publications have been aggregated in the disciplinary areas to which their authors belong. This is a new and novel approach in this field of study[12], especially considering the breadth and diversity of the scientific production under investigation: for the period under study there were over 36,000 Italian university research staff in technical-scientific disciplines ("hard-sciences"), with 4,000 presenting problems of homonymy in their last name and initial. Tables 3 and 4 show the respective descriptive statistics for the input and output variables of the model, classified by disciplinary area.

| UDA | Active universities | FP | AP | RF | PU | PC | SS |
|---|---|---|---|---|---|---|---|
| 1 | 53 | 937 | 1092 | 1078 | 3034 | 2805 | 3526 |
| 2 | 49 | 795 | 934 | 787 | 8361 | 7620 | 21843 |
| 3 | 50 | 942 | 1183 | 1025 | 12347 | 11036 | 30955 |
| 4 | 42 | 385 | 484 | 422 | 1706 | 1530 | 3339 |
| 5 | 52 | 1429 | 1588 | 1848 | 12770 | 10504 | 44359 |
| 6 | 44 | 2448 | 3327 | 4796 | 23766 | 21346 | 81797 |
| 7 | 34 | 956 | 892 | 1117 | 3006 | 2564 | 5597 |
| 8 | 39 | 1011 | 1185 | 1349 | 1136 | 972 | 1698 |
| 9 | 50 | 1548 | 1446 | 1356 | 6057 | 5441 | 8870 |

***Table 2: Average input and output values of the Italian university system, period 2001-2003***

| | | FP | | | | AP | | | | RF | | | |
|---|---|---|---|---|---|---|---|---|---|---|---|---|---|
| UDA | Universities Surveyed | Ave | Min | Max | Std Dev | Ave | Min | Max | Std Dev | Ave | Min | Max | Std Dev |
| 1 | 53 | 18 | 0 | 86 | 18 | 21 | 1 | 109 | 20 | 21 | 2 | 69 | 17 |
| 2 | 49 | 16 | 1 | 72 | 15 | 19 | 0 | 62 | 17 | 16 | 1 | 52 | 12 |
| 3 | 50 | 20 | 0 | 69 | 19 | 25 | 1 | 103 | 22 | 22 | 2 | 81 | 19 |
| 4 | 42 | 10 | 1 | 33 | 8 | 12 | 1 | 37 | 9 | 11 | 1 | 32 | 7 |
| 5 | 52 | 27 | 0 | 103 | 24 | 30 | 0 | 117 | 28 | 35 | 0 | 154 | 31 |
| 6 | 44 | 57 | 0 | 226 | 45 | 77 | 1 | 385 | 72 | 111 | 1 | 688 | 128 |
| 7 | 34 | 29 | 1 | 86 | 26 | 27 | 0 | 80 | 23 | 34 | 1 | 112 | 30 |
| 8 | 39 | 27 | 1 | 135 | 31 | 32 | 2 | 134 | 33 | 36 | 2 | 149 | 38 |
| 9 | 50 | 33 | 1 | 169 | 40 | 31 | 1 | 161 | 36 | 29 | 0 | 136 | 30 |

***Table 3: Statistics for inputs used in the model (averages of values from 2001 to 2003)***

| | PU | | | | PC | | | | SS | | | |
|---|---|---|---|---|---|---|---|---|---|---|---|---|
| UDA | Ave | Min | Max | Std Dev | Ave | Min | Max | Std Dev | Ave | Min | Max | Std Dev |
| 1 | 21 | 1 | 94 | 19 | 13 | 1 | 55 | 11 | 24 | 0 | 110 | 24 |
| 2 | 56 | 1 | 260 | 50 | 19 | 0 | 86 | 17 | 150 | 4 | 724 | 147 |
| 3 | 87 | 4 | 367 | 79 | 49 | 2 | 226 | 46 | 220 | 10 | 995 | 213 |
| 4 | 15 | 0 | 41 | 12 | 7 | 0 | 22 | 6 | 29 | 0 | 101 | 25 |
| 5 | 79 | 2 | 317 | 69 | 42 | 1 | 175 | 38 | 275 | 5 | 1120 | 254 |
| 6 | 166 | 2 | 653 | 144 | 87 | 1 | 328 | 76 | 578 | 5 | 2230 | 507 |
| 7 | 30 | 0 | 116 | 30 | 17 | 0 | 76 | 19 | 56 | 0 | 226 | 60 |
| 8 | 11 | 1 | 45 | 10 | 7 | 1 | 29 | 6 | 16 | 1 | 55 | 15 |
| 9 | 45 | 0 | 203 | 45 | 28 | 0 | 131 | 29 | 66 | 0 | 295 | 68 |

***Table 4: Statistics for outputs used in the model (averages of values from 2001 to 2003)***

## 3. Technical and cost efficiency: analysis by disciplinary area

DEA simulations used to calculate efficiency scores, which equal the radial distance of each DMU from the efficient frontier, were carried out with the following hypotheses:

i. Constant returns to scale: both literature analysis and tests conducted by the current authors lead to rejection of the hypothesis that returns vary with the scale of the DMU. In addition, although the object of investigation is the individual university disciplinary area, their internal research groups are generally organized with a functional maximum of 7 or 8 professionals, which is an order of magnitude below the size of UDAs.

ii. Input orientation: the score represents the maximum equi-proportional decrease in all inputs (outputs remaining equal). This model emphasizes the management objective of reducing input resources while maintaining equal output. A score value of 1 indicates fully efficient units.

Table 5 shows descriptive statistics for efficiency scores, classified by disciplinary area[13]. Average technical efficiency varies from a minimum of 0.364 for disciplinary area 8 (civil engineering and architecture[14]) up to a maximum of 0.624 for disciplinary area 5 (biology). Allocative efficiency is consistently higher, with the exception of UDA 3 (chemistry) where average allocative efficiency is only 0.477 compared to an average technical efficiency of 0.487. The chemistry area also proved to be the least efficient of all, with an average cost efficiency score of 0.226.

| | UDA | 1 | 2 | 3 | 4 | 5 | 6 | 7 | 8 | 9 |
|---|---|---|---|---|---|---|---|---|---|---|
| | Number of universities | 53 | 49 | 50 | 42 | 52 | 44 | 34 | 39 | 50 |
| TE | *Ave* | 0.573 | 0.536 | 0.487 | 0.606 | 0.624 | 0.462 | 0.529 | 0.364 | 0.531 |
| | *Min* | 0.251 | 0.09 | 0.079 | 0.083 | 0.239 | 0.204 | 0.139 | 0.017 | 0.04 |
| | *Std Dev* | 0.220 | 0.210 | 0.174 | 0.249 | 0.199 | 0.188 | 0.271 | 0.274 | 0.268 |
| AE | *Ave* | 0.732 | 0.876 | 0.477 | 0.816 | 0.661 | 0.791 | 0.843 | 0.783 | 0.853 |
| | *Min* | 0.519 | 0.472 | 0.222 | 0.452 | 0.474 | 0.574 | 0.514 | 0.433 | 0.471 |
| | *Std Dev* | 0.111 | 0.084 | 0.098 | 0.146 | 0.092 | 0.092 | 0.104 | 0.124 | 0.096 |
| CE | *Ave* | 0.422 | 0.476 | 0.226 | 0.499 | 0.416 | 0.364 | 0.441 | 0.277 | 0.446 |
| | *Min* | 0.182 | 0.043 | 0.062 | 0.045 | 0.15 | 0.156 | 0.119 | 0.014 | 0.03 |
| | *Std Dev* | 0.197 | 0.201 | 0.080 | 0.227 | 0.168 | 0.156 | 0.232 | 0.210 | 0.219 |

***Table 5: Values for technical efficiency (TE), allocative efficiency (AE) and cost efficiency (CE), by disciplinary area***

The data shown in Table 6 bring out the deviations in average scores for technical efficiency between clusters of research institutions. For instance, within UDA 1 (mathematics), 6 universities lie at the frontier of technical efficiency. The scores for the remaining "inefficient" institutions were subdivided into tertile groups. The first tertile has an average score of 0.714, while the second tertile averages 0.482 and the third tertile averages 0.345. If UDA 8 is excluded (the disciplinary area with the particular characteristics noted above), the average deviation of efficiency scores between the first and second tertiles is 28%, rising to an average deviation of 54% between the first and third tertiles. Table 7 shows the results for cost-efficiency scores subjected to the same type of analysis. The number of efficient universities per disciplinary area never rises above 2. Deviations in efficiency among clusters of universities are very evident. The data concerning the last tertile are particularly notable: it accounts for about 30% of all universities and has an average cost efficiency of 22%. This means that 3 out of 10 universities would have a margin for improvement sufficiently high that they could achieve the same output while employing only one fifth of the economic resources presently at their disposal.

| UDA | Efficient universities | Inefficient universities | | |
|---|---|---|---|---|
| | | 1st tertile | 2nd tertile | 3rd tertile |
| 1 | 6 (out of 53) | 0.714 | 0.485 | 0.345 |
| 2 | 2 (out of 49) | 0.721 | 0.496 | 0.321 |
| 3 | 1 (out of 50) | 0.639 | 0.458 | 0.323 |
| 4 | 5 (out of 42) | 0.777 | 0.559 | 0.302 |
| 5 | 3 (out of 52) | 0.787 | 0.614 | 0.391 |
| 6 | 3 (out of 44) | 0.548 | 0.432 | 0.277 |
| 7 | 6 (out of 34) | 0.604 | 0.431 | 0.267 |
| 8 | 2 (out of 39) | 0.609 | 0.273 | 0.123 |
| 9 | 8 (out of 50) | 0.651 | 0.426 | 0.249 |

***Table 6: Average values of technical efficiency for clusters of universities***

| UDA | Efficient universities | Inefficient universities | | |
|---|---|---|---|---|
| | | 1st tertile | 2nd tertile | 3rd tertile |
| 1 | 2 (out of 53) | 0.581 | 0.365 | 0.253 |
| 2 | 2 (out of 49) | 0.641 | 0.432 | 0.277 |
| 3 | 0 (out of 50) | 0.306 | 0.220 | 0.147 |
| 4 | 1 (out of 42) | 0.714 | 0.490 | 0.238 |
| 5 | 2 (out of 52) | 0.517 | 0.394 | 0.258 |
| 6 | 1 (out of 44) | 0.476 | 0.344 | 0.219 |
| 7 | 2 (out of 34) | 0.605 | 0.369 | 0.228 |
| 8 | 1 (out of 39) | 0.452 | 0.216 | 0.094 |
| 9 | 2 (out of 50) | 0.637 | 0.399 | 0.233 |

***Table 7: Average values of cost efficiency for clusters of universities***

## 4. University ranking

In each disciplinary area, university scores and rankings have been calculated according to the following efficiency indicators: technical efficiency, allocative efficiency, cost efficiency, output to number of scientists ratio, and output to costs ratio.

Table 8 provides the example of scores and rankings for the full series of efficiency indicators for all universities operating in mathematics.

Five of the six universities identified as technically efficient have less than 20 research staff. In spite of their small staff size, these universities are also among the leaders of rankings for the output per scientist indicator. The largest university (261 employees) has a technical efficiency score of 0.559, thus falling in the 56$^{th}$ percentile.

Allocative efficiency scores have a normal distribution with a rather high average (0.73) and a very low coefficient of variation (0.15). It follows logically that cost efficiency depends largely on technical efficiency rather than on allocative efficiency.

In general, the table shows a strong correlation of both technical efficiency and cost efficiency scores with the simple ratio calculations of output per scientist (publications per scientist) and output per cost unit (publications per €100,000 of staff cost). However, this does not mean that the two measurement models (DEA and single output/single input ratios) produce exactly the same rankings. This point is further discussed in the following paragraphs.

| DMU | Scientists | TE score | TE rank | AE score | AE rank | CE score | CE rank | Output/Scientist score | Output/Scientist rank | Output/Cost score | Output/Cost rank |
|---|---|---|---|---|---|---|---|---|---|---|---|
| University 1 | 4 | 0.251 | 53 | 0.819 | 10 | 0.206 | 51 | 0.182 | 51 | 0.145 | 50 |
| University 2 | 5 | 0.471 | 30 | 0.716 | 31 | 0.338 | 31 | 0.333 | 27 | 0.244 | 26 |
| University 3 | 6 | 0.749 | 14 | 0.631 | 44 | 0.473 | 14 | 0.444 | 14 | 0.342 | 11 |
| University 4 | 7 | 1 | 1 | 0.723 | 27 | 0.723 | 5 | 0.600 | 6 | 0.523 | 4 |
| University 5 | 7 | 0.364 | 45 | 0.652 | 42 | 0.237 | 47 | 0.200 | 49 | 0.171 | 45 |
| University 6 | 7 | 0.384 | 42 | 0.522 | 52 | 0.201 | 52 | 0.136 | 53 | 0.103 | 53 |
| University 7 | 8 | 0.659 | 15 | 0.552 | 50 | 0.363 | 26 | 0.333 | 27 | 0.263 | 22 |
| University 8 | 13 | 1 | 1 | 1 | 1 | 1 | 1 | 0.974 | 1 | 0.640 | 3 |
| University 9 | 14 | 1 | 1 | 1 | 1 | 1 | 1 | 0.952 | 2 | 0.723 | 1 |
| University 10 | 15 | 0.545 | 25 | 0.764 | 19 | 0.417 | 22 | 0.386 | 21 | 0.301 | 19 |
| University 11 | 15 | 0.261 | 51 | 0.908 | 4 | 0.237 | 47 | 0.244 | 43 | 0.171 | 46 |
| University 12 | 16 | 0.544 | 26 | 0.532 | 51 | 0.29 | 40 | 0.229 | 45 | 0.183 | 43 |
| University 13 | 18 | 1 | 1 | 0.921 | 3 | 0.921 | 3 | 0.849 | 3 | 0.640 | 2 |
| University 14 | 19 | 0.646 | 17 | 0.725 | 26 | 0.468 | 15 | 0.429 | 17 | 0.326 | 13 |
| University 15 | 19 | 1 | 1 | 0.578 | 49 | 0.578 | 9 | 0.439 | 15 | 0.320 | 16 |
| University 16 | 20 | 0.756 | 13 | 0.798 | 13 | 0.603 | 8 | 0.574 | 8 | 0.436 | 8 |
| University 17 | 22 | 0.51 | 28 | 0.519 | 53 | 0.265 | 44 | 0.224 | 47 | 0.166 | 48 |
| University 18 | 23 | 0.389 | 41 | 0.878 | 6 | 0.341 | 30 | 0.279 | 38 | 0.207 | 38 |
| University 19 | 26 | 0.775 | 12 | 0.579 | 48 | 0.449 | 17 | 0.462 | 12 | 0.325 | 14 |
| University 20 | 28 | 0.45 | 35 | 0.721 | 29 | 0.324 | 36 | 0.313 | 30 | 0.233 | 30 |
| University 21 | 28 | 0.806 | 9 | 0.808 | 12 | 0.652 | 6 | 0.576 | 7 | 0.471 | 6 |
| University 22 | 32 | 0.911 | 7 | 0.684 | 38 | 0.623 | 7 | 0.632 | 5 | 0.451 | 7 |
| University 23 | 36 | 0.587 | 22 | 0.759 | 21 | 0.446 | 20 | 0.449 | 13 | 0.323 | 15 |
| University 24 | 37 | 0.364 | 45 | 0.795 | 15 | 0.289 | 41 | 0.279 | 40 | 0.203 | 39 |
| University 25 | 38 | 0.253 | 52 | 0.718 | 30 | 0.182 | 53 | 0.174 | 52 | 0.129 | 52 |
| University 26 | 42 | 1 | 1 | 0.883 | 5 | 0.883 | 4 | 0.744 | 4 | 0.516 | 5 |
| University 27 | 45 | 0.648 | 16 | 0.596 | 47 | 0.386 | 25 | 0.296 | 31 | 0.224 | 31 |
| University 28 | 47 | 0.335 | 49 | 0.676 | 39 | 0.227 | 49 | 0.221 | 48 | 0.164 | 49 |
| University 29 | 47 | 0.408 | 39 | 0.692 | 37 | 0.282 | 42 | 0.284 | 37 | 0.198 | 40 |
| University 30 | 49 | 0.488 | 29 | 0.858 | 8 | 0.419 | 21 | 0.401 | 20 | 0.287 | 21 |
| University 31 | 53 | 0.381 | 44 | 0.798 | 13 | 0.304 | 38 | 0.296 | 32 | 0.220 | 32 |
| University 32 | 54 | 0.319 | 50 | 0.838 | 9 | 0.267 | 43 | 0.264 | 41 | 0.185 | 42 |
| University 33 | 55 | 0.407 | 40 | 0.817 | 11 | 0.332 | 32 | 0.333 | 27 | 0.240 | 29 |
| University 34 | 55 | 0.788 | 11 | 0.701 | 35 | 0.553 | 11 | 0.470 | 11 | 0.328 | 12 |
| University 35 | 56 | 0.344 | 47 | 0.709 | 32 | 0.244 | 46 | 0.228 | 46 | 0.168 | 47 |
| University 36 | 60 | 0.453 | 34 | 0.723 | 27 | 0.328 | 35 | 0.279 | 39 | 0.217 | 34 |
| University 37 | 67 | 0.81 | 8 | 0.705 | 33 | 0.571 | 10 | 0.554 | 10 | 0.413 | 9 |
| University 38 | 70 | 0.639 | 18 | 0.699 | 36 | 0.447 | 19 | 0.414 | 18 | 0.312 | 17 |
| University 39 | 80 | 0.604 | 21 | 0.742 | 23 | 0.448 | 18 | 0.413 | 19 | 0.301 | 20 |
| University 40 | 81 | 0.559 | 23 | 0.705 | 33 | 0.394 | 24 | 0.346 | 24 | 0.247 | 25 |
| University 41 | 84 | 0.428 | 37 | 0.772 | 16 | 0.331 | 33 | 0.289 | 36 | 0.218 | 33 |
| University 42 | 90 | 0.336 | 48 | 0.661 | 41 | 0.222 | 50 | 0.196 | 50 | 0.141 | 51 |
| University 43 | 96 | 0.634 | 19 | 0.86 | 7 | 0.545 | 12 | 0.564 | 9 | 0.385 | 10 |
| University 44 | 100 | 0.439 | 36 | 0.667 | 40 | 0.293 | 39 | 0.290 | 34 | 0.211 | 36 |
| University 45 | 121 | 0.518 | 27 | 0.772 | 16 | 0.4 | 23 | 0.371 | 22 | 0.262 | 23 |
| University 46 | 127 | 0.455 | 33 | 0.726 | 25 | 0.331 | 33 | 0.292 | 33 | 0.210 | 37 |
| University 47 | 129 | 0.469 | 31 | 0.766 | 18 | 0.359 | 27 | 0.342 | 25 | 0.244 | 27 |
| University 48 | 129 | 0.802 | 10 | 0.613 | 46 | 0.491 | 13 | 0.289 | 35 | 0.217 | 35 |
| University 49 | 145 | 0.421 | 38 | 0.728 | 24 | 0.307 | 37 | 0.260 | 42 | 0.195 | 41 |
| University 50 | 151 | 0.61 | 20 | 0.745 | 22 | 0.454 | 16 | 0.430 | 16 | 0.305 | 18 |
| University 51 | 164 | 0.457 | 32 | 0.763 | 20 | 0.349 | 29 | 0.341 | 26 | 0.242 | 28 |
| University 52 | 178 | 0.384 | 42 | 0.637 | 43 | 0.245 | 45 | 0.234 | 44 | 0.177 | 44 |
| University 53 | 261 | 0.559 | 23 | 0.629 | 45 | 0.352 | 28 | 0.361 | 23 | 0.254 | 24 |

***Table 8: Efficiency scores and associated rankings for universities carrying out research in mathematics***

The DEA efficiency scores and partial indicators permit analysis of the performance of individual areas within the same university. Table 9 shows the data for the example of one Italian university. As can be noted, the university is active in all disciplinary areas but its performance varies from discipline to discipline. This particular university showed the best technical efficiency in disciplinary area 6, medicine, with a rank of sixth at the national level. The other measurement models for medicine give almost the same placement of this university for this area. However the models produce different results for area 4 (earth sciences): the DEA technical efficiency score brings in a national rank of sixteenth for this university, against a rank of eighth when using the output/scientist indicator. In this area, the specificities of the indicators in question have a clear impact on the ranking. The same is true for the indicators in area 5 (biology): the university rates quite well (17th) in DEA ranking for technical efficiency, which outshines its rankings in publications per scientist (27$^{th}$) and publications per cost unit (29$^{th}$).

| UDA | Total employees | Total publications | TE | | AE | | CE | | Output/scientist | | Output/cost | |
|---|---|---|---|---|---|---|---|---|---|---|---|---|
| | | | score | rank | score | rank | score | rank | score | rank | score | rank |
| 1 | 164 | 56 | 0.457 | 32 | 0.763 | 20 | 0.349 | 29 | 0.341 | 26 | 0.242 | 28 |
| 2 | 92 | 117 | 0.550 | 19 | 0.943 | 9 | 0.518 | 15 | 1.264 | 16 | 0.910 | 17 |
| 3 | 122 | 148 | 0.416 | 34 | 0.534 | 10 | 0.222 | 23 | 1.210 | 30 | 0.874 | 34 |
| 4 | 53 | 33 | 0.696 | 16 | 0.940 | 10 | 0.654 | 10 | 0.633 | 8 | 0.458 | 6 |
| 5 | 168 | 150 | 0.719 | 17 | 0.617 | 38 | 0.444 | 18 | 0.895 | 27 | 0.661 | 29 |
| 6 | 267 | 263 | 0.573 | 6 | 0.840 | 12 | 0.481 | 5 | 0.986 | 9 | 0.753 | 7 |
| 7 | 164 | 53 | 0.374 | 24 | 0.950 | 5 | 0.356 | 19 | 0.325 | 22 | 0.240 | 22 |
| 8 | 47 | 8 | 0.336 | 15 | 0.883 | 7 | 0.297 | 12 | 0.177 | 14 | 0.130 | 14 |
| 9 | 217 | 93 | 0.380 | 32 | 0.914 | 11 | 0.348 | 32 | 0.428 | 35 | 0.295 | 35 |

***Table 9: Efficiency scores and associated rankings for the disciplinary areas (1 to 9) of a sample University***

The above results and analysis invited further investigations and a broader scale of comparison between results from the DEA model being proposed and from the more simplified type of single-output/single-input model. Table 10 presents statistics describing the distribution of variations in efficiency rankings for all the universities active in each disciplinary area. The table compares rankings from technical efficiency analysis with those taken from the ratio of average number of publications per scientist, and shows substantial variations in ranking in all disciplinary areas, with differences occurring in at least two thirds of the DMUs surveyed. The average difference in ranking obtained from the two models varies from a minimum of 2 positions per UDAs 2 and 7 (agriculture and veterinary sciences; and civil engineering and architecture), up to a maximum of 8, recorded for UDA 5 (biology). The extreme variation in some areas (in particular the third, chemistry), shows that ranking shifts can be truly remarkable: one DMU in this area drops 33 positions if the ranking is taken from its publications per scientist, rather from its technical efficiency score. It follows that choosing an appropriate model of productivity measurement is a critical task in planning and implementing an evaluation system. Indeed, despite the strong overall correlation between scores obtained from different measurement models, the models can induce notably different rankings for the individual universities. If the evaluation system is being devised for resource allocation purposes then the parameters to be used must be targeted with extreme care.

| UDA | Variations | Max variation | Average variation | Median | Coefficient of variation |
|---|---|---|---|---|---|
| 1 | 44 (out of 53) | 25 | 5 | 3 | 1.082 |
| 2 | 42 (out of 49) | 12 | 3 | 3 | 0.885 |
| 3 | 44 (out of 50) | 33 | 7 | 4 | 1.051 |
| 4 | 37 (out of 42) | 18 | 4 | 3,5 | 0.911 |
| 5 | 41 (out of 52) | 23 | 8 | 8 | 0.734 |
| 6 | 38 (out of 44) | 22 | 5 | 3,5 | 1.068 |
| 7 | 23 (out of 34) | 7 | 2 | 2 | 0.965 |
| 8 | 26 (out of 39) | 7 | 2 | 1,5 | 0.872 |
| 9 | 36 (out of 50) | 14 | 4 | 2 | 1.003 |

***Table 10: Variations in ranking of university discipline areas as obtained from technical efficiency scores and output/scientist ratios***

## 5. Conclusions

The comparative analysis of methodologies being used for research productivity measurements in universities and public research institutions call for careful study. This reflection becomes essential if national policy makers intend to use them for allocation purposes, as is currently the situation in some countries, including Italy. In particular, it is obligatory to establish whether scarce resources should be allocated to universities according to a relatively fixed idea of excellence, implied by defining universal and unvarying algorithms and measurement methods, or according to more tailored sets of strategic criteria, varying with location and time. Although the question seems open and relevant, it is evident that a number of national governments already favour an “outcome control” evaluation that measures research performance, sometimes to guide their allocation decisions. In particular, governments generally prefer peer-review methodologies over other assessment methods, although they present complexities in implementation and have truly notable costs and time requirements. Moreover, peer-review methodologies, taken alone, lend themselves poorly to measuring productivity. Bibliometric methods, on the other hand, are low-cost, non-invasive, easy to implement, with the possibility of rapid updates and inter-temporal comparison. They are based on objective qualitative-quantitative data, have a high degree of representativeness of the surveyed universe and allow international comparisons. However, the bibliometric approach has so far only been used only as a tool for analysis of clearly-delimited investigation areas, for thematic surveys and other specific studies, largely because of the methodological limitations that the technique has embodied.

This study focuses on the above-mentioned limitations and an attempt to overcome them. A proposal is made for a cost-efficiency measurement system for an entire academic research system based on non-parametric techniques of bibliometric assessment (data envelopment analysis). This technique offers a number of novel advantages compared to the current state of the art, namely:

- the possibility of giving due consideration to the various forms of input (and their costs) and outputs for the research laboratories under survey; and to discriminate the two single components of efficiency: technical and allocative;
- the possibility of giving consideration to the heterogeneity of universities in terms of their resource mixes and the specificities of their research disciplines;
- above all, the possibility of identifying output levels with accuracy never attained

before in large-scale studies in the literature, through the method of linking scientific publications with each individual research author and subsequently aggregating the authors and publications by disciplinary areas.

The results typically show much variation in the rankings of the disciplinary areas within and across universities, depending on the efficiency indicator employed. This should be carefully considered when formulating research evaluation exercises, at any level.

The methodological assumptions and consequent implications induce a certain level of caution in interpreting some of the results and suggest further investigation and fine tuning. The authors will follow up on these issues, in particular substituting the IF values with citation counts, enlarging the field of observation to include more scientific journals (from the CD-rom version of SCI$^{TM}$ to WoS) and expanding the time period of analysis. The aim is to providea bibliometric model that can be a robust benchmarking tool for research management and a valid integrating instrument in peer-reviews to support policy makers' decisions.

---

**Notes**

[1] The main examples are: the Science Citation Index (SCI™) for technical-scientific disciplines, the Social Science Citation Index for social sciences, and the Arts and Humanities Citation Index for humanities, which are part of the Web of Science. See http://scientific.thomsonreuters.com

[2] This is particularly true for the Italian research system where the number of patents filed by universities is extremely low and the publications rate is inversely high compared to other countries (Abramo, 2007).

[3] www.rae.ac.uk

[4] The Italian university system adopts a scientific classification system comprising 14 “disciplinary areas”, which in turn include 370 "scientific disciplinary sectors". For details see http://www.miur.it/userfiles/115.htm. Each research scientist belongs to only one disciplinary sector. Due to their high specialization level, each disciplinary sector may be considered homogeneous for the purposes of research productivity assessment.

[5] All articles indexed by Thomson Reuters’ WoS are classified into 245 categories. WoS categories differ from the scientific disciplinary sectors (370) of the Italian academic system.

[6] Mathematics and computer sciences; physics; chemistry; earth sciences; biology; medicine; agriculture and veterinary sciences; industrial and information engineering.; civil engineering and architecture.

[7] https://dalia.cineca.it/php4/inizio_access_cnvsu.php

[8] The weight of assistant professors equals “1”, while the weights of associate professors and full professors were calculated normalizing the wage costs for these categories to that of assistant professors. Note that professors wages are centrally determined by rank and seniority and do not vary from one university to another. For each rank we have considered the seniority-average wage.

[9] This indicator represents a proxy measure of the total number of citations traceable to the scientific production of the unit under consideration. The authors are aware of the intrinsic limitations of such approximation, as well as of the recommendations contained in the literature on this issue (Moed and Van Leeuwen, 1996, Weingart, 2004). However, as they did not have access to data on the actual number of citations and as their purpose was not that of providing a ranking of the surveyed units, but only to present a methodology, the authors decided to proceed with the study on the basis of this proxy measures. It should be noted that the impact factors can be readily replaced with actual figures on citations at a later stage, when available.

[10] The distribution of impact factors of journals differs substantially from one scientific category to another. The normalization of each journal’s impact factor with respect to the category average permits limiting the distortions in comparing performances between different categories.

[11] While representing measures of total productivity, efficiency scores derive exclusively from the choice of input and output variables and thus depend on the simplifying hypotheses adopted for the analysis. Possible distortions might be due to the following: incongruities in the allocation of input time between research and teaching or between different types of research (basic/applied), or the uneven distribution of input factors such as laboratories, libraries, scientific instruments, materials, temporary staff and PhD students; varying tendencies of researchers to produce their output in alternative forms of publication, or uneven agglomeration and scope economies.

[12] To the authors' knowledge, only two studies on the disambiguation of scientific publications have been published. Wooding et al (2006) utilise a recursive algorithm for disambiguating publications in the field of arthritic diseases. Torvik et al (2005) apply stochastic similarity metrics to publications listed in the Medline© directory of the American National Library of Medicine.

[13] The rankings resulting from the identification of the frontier and the calculation of efficiency scores do not include particularly small disciplinary areas. A minimum threshold was identified by means of an empirical recursive procedure, such that research units with less than 5 staff members were excluded. The choice of this threshold was due to the empirical observation that the values of total factor productivity might be distorted by cases where input values were all close to zero.

[14] It should be noted that this research area falls on the boundary between scientific-technological disciplines (civil engineering) and arts (architecture). Output data cannot be surveyed to an exhaustive extent using the SCI database alone, therefore the interpretation of the data recorded within this area requires a certain degree of caution.